\documentclass[fleqn,twoside]{article}
\usepackage{espcrc2}
\usepackage{latexsym}
\usepackage{euscript}

\usepackage{graphicx}
\usepackage[figuresright]{rotating}

\def\simge{\mathrel{%
   \rlap{\raise 0.511ex \hbox{$>$}}{\lower 0.511ex \hbox{$\sim$}}}}   
\def\simle{\mathrel{   
   \rlap{\raise 0.511ex \hbox{$<$}}{\lower 0.511ex \hbox{$\sim$}}}}   
\def\slashchar#1{\setbox0=hbox{$#1$}           
   \dimen0=\wd0                                 
   \setbox1=\hbox{/} \dimen1=\wd1               
   \ifdim\dimen0>\dimen1                        
      \rlap{\hbox to \dimen0{\hfil/\hfil}}      
      #1                                        
   \else                                        
      \rlap{\hbox to \dimen1{\hfil$#1$\hfil}}   
      /                                         
   \fi}                                         %

\def\simge{\mathrel{%
   \rlap{\raise 0.511ex \hbox{$>$}}{\lower 0.511ex \hbox{$\sim$}}}}   
\def\simle{\mathrel{   
   \rlap{\raise 0.511ex \hbox{$<$}}{\lower 0.511ex \hbox{$\sim$}}}}   
\def\slashchar#1{\setbox0=\hbox{$#1$}           
   \dimen0=\wd0                                 
   \setbox1=\hbox{/} \dimen1=\wd1               
   \ifdim\dimen0>\dimen1                        
      \rlap{\hbox to \dimen0{\hfil/\hfil}}      
      #1                                        
   \else                                        
      \rlap{\hbox to \dimen1{\hfil$#1$\hfil}}   
      /                                         
   \fi}

\newcommand{\AmS}{{\protect\the\textfont2
  A\kern-.1667em\lower.5ex\hbox{M}\kern-.125emS}}

\newcommand{\ba}{\begin{equation} \left\{ \begin{array}{lr}}
\newcommand{\ea}{\end{array} \right. \end{equation}}

\newcommand{\bea}{\begin{eqnarray}}
\newcommand{\eea}{\end{eqnarray}}
\newcommand{\be}{\begin{equation}}
\newcommand{\ee}{\end{equation}}

\title{\vspace{-4.8cm}
       \rightline{\normalsize ROM2F/2004/20}
       \vspace{-0.1cm}
       \rightline{\normalsize DESY 04-144}
       \vspace{-0.1cm}
       \rightline{\normalsize FTUAM-04-19}
       \vspace{-0.1cm}
       \rightline{\normalsize IFT UAM-CSIC/04-45}
       \vspace{-0.1cm}
       \rightline{\normalsize August 2004}
       \vspace{2.0cm}
NLO anomalous dimension of multiplicatively renormalizable
  four--fermion operators in Schr\"odinger Functional schemes\thanks{Talk
  given at Lattice 2004 by F.P.}}

\author{Filippo Palombi\address[Fermi]{``E.~Fermi''~Research Center,
    c/o Compendio Viminale --  pal.~F, I-00184 Rome, Italy}, Carlos
  Pena\address[DESY]{DESY, Theory Group, Notkestra\ss e 85, D-22603
    Hamburg, Germany} and Stefan Sint\address[AUTONOMA]{Departamento de F\'{\i}sica Te\'orica C-XI and
  Instituto de F\'{\i}sica Te\'orica C-XVI, Universidad Aut\'onoma de
  Madrid, Cantoblanco, E-28049 Madrid, Spain} \ \ (ALPHA Collaboration) }
       
\begin{document}

\begin{abstract}

Renormalization constants for multiplicatively renormalizable parity-odd four--fermion operators
are computed in various different Schr\"odinger Functional (SF) schemes and lattice regularizations
with Wilson quarks at one--loop order in perturbation theory. Our results are used in
the calculation of their NLO anomalous dimensions, through matching to continuum schemes.
They also enable a comparison of the two--loop perturbative RG running to the previously
obtained nonperturbative one in the region of small renormalized
coupling.

\end{abstract}

\maketitle

\section{Introduction}

Four--fermion operators occur in the transition amplitudes of many physical processes within
the SM and beyond. An example is provided by the oscillations of the $K^0 - \bar
K^0$ system, where the mixing amplitude is given by a matrix element
of the form
\be
\langle \bar K^0|{\cal O}^+_{\rm VV+AA}|K^0\rangle = \frac{8}{3}F_K^2m_K^2B_K.
\ee
Renormalizing the operator ${\cal O}^+_{\rm VV+AA}$ on the lattice
with Wilson quarks requires special care, as it mixes with other
four--fermion operators. The situation is simpler in the framework of tmQCD,
where the operator in question can be mapped onto its parity-odd counterpart
${\cal O}^+_{\rm VA+AV}$, which is multiplicatively renormalizable
\cite{Frezzotti:2000nk}. Nonperturbative renormalization of the 
latter in SF schemes has been reported in \cite{Guagnelli:2002rw},
where its scale dependence has been computed in the continuum limit
through finite size scaling techniques. On the other hand, the corresponding one--loop perturbative
calculation is required in order to make contact to continuum schemes.
It can also be used for an analytical estimation of the perturbative lattice artefacts.

\section{SF renormalization}

Renormalization of a local fermion operator is pursued in a SF scheme
by imposing a suitable renormalization condition at vanishing
renormalized quark mass on a correlation
function, properly defined to probe the operator by bilinear boundary
quark sources
\begin{eqnarray}
S_{f_1f_2}[\Gamma] & = & a^6\sum_{\bf y,z} \ \bar\zeta_{f_1}
({\bf y})\ \Gamma \ \zeta_{f_2}({\bf z}), \cr \cr
S'_{f_1f_2}[\Gamma] & = & a^6\sum_{\bf y',z'} \ \bar\zeta_{f_1}'
({\bf y'})\ \Gamma\ \zeta_{f_2}'({\bf z'}).
\end{eqnarray}
Here the boundary fields $\zeta$ and $\bar\zeta$ are defined at
$x_0=0$, while $\zeta'$ and $\bar\zeta'$ are defined at $x_0=T$;
$\Gamma$ is a generic Dirac matrix and $f_i$ are flavour indices. The
simplest way to probe the four--fermion operator
\begin{eqnarray}
{\cal O}_{\rm VA+AV}^\pm & = &
\frac{1}{2}\bigl[\bar\psi_1\gamma_\mu\psi_2\bar\psi_3\gamma_\mu\gamma_5\psi_4
  + \cr 
  &  & \ \ \ \bar\psi_1\gamma_\mu\gamma_5\psi_2\bar\psi_3\gamma_\mu\psi_4\pm
  (2\leftrightarrow 4)\bigr]
\end{eqnarray}
is by inserting it in a correlation function with three bilinear boundary
sources
\be
h^\pm(x_0) = \frac{1}{L^3}\langle S'_{53}[\Gamma_3]{\cal O}^\pm(x)S_{21}[\Gamma_1]S_{45}[\Gamma_2]\rangle.
\ee 
Parity conservation implies that one has to provide a parity--odd
choice of the set $[\Gamma_1,\Gamma_2,\Gamma_3]$ in order to get a
nonvanishing correlation function. Provided this, the ultraviolet
divergences of the boundary fields must be removed in order to isolate
the divergences of the operator. To this purpose $h^\pm(x_0)$ can be
normalized as
\be
\tilde h^\pm(x_0) = \frac{h^\pm(x_0)}{f_1^\alpha\ 
  k_1^\beta}\biggr|_{\alpha+\beta = 3/2},
\ee
where we have introduced the standard boundary correlation functions with no operator
insertion
\begin{eqnarray}
f_1 & = & -\frac{1}{\ \ L^6}\langle S'[\gamma_5]S[\gamma_5]\rangle, \cr
k_1 & = & -\frac{1}{3L^6}\sum_{k=1}^3\langle S'[\gamma_k]S[\gamma_k]\rangle.
\end{eqnarray}
The condition $\alpha + \beta = 3/2$ makes $\tilde h^\pm(x_0)$ free of
boundary divergences and suitable to be renormalized through a
multiplicative condition
\be 
\tilde h_R^\pm(T/2) = \tilde h^\pm_{\rm tree}(T/2) \ \ {\rm at}\ m_R=0,
\ee
where the subtraction point is at $\mu=1/L$, $m_R$ is the
renormalized quark mass, the operator ${\cal O}^\pm_{\rm VA+AV}$
is placed in the middle of the time extent of the SF and
\be
\tilde h_R^\pm(T/2) = Z^{\pm}_{\rm VA+AV}(g_0,a/L)\ \tilde h^\pm(T/2).
\ee
Apart from the indicated dependence, the renormalization constant
$Z_{\rm VA+AV}^\pm$ depends upon every detail, such as the value of \
$T/L$ (in this work $T/L=1$), the choice of the Dirac structure of the
sources and the exponents $\alpha,\beta$ in eq.~(5). A number of possible combinations exist,
each one generating a different renormalization scheme in the
framework of the SF. In Table~1 we report the ones we have considered.

\begin{table}[t]
 \centering
 \begin{tabular}{crcc}
 \hline\\[-1.8ex]
 scheme & $[\Gamma_1,\Gamma_2,\Gamma_3]$\hskip 0.8cm\  & $\alpha$ & $\beta$ \\
 \hline\\[-1.8ex]
 I    & $[\gamma_5,\gamma_5,\gamma_5]$ & $3/2$ & $0$ \cr 

 II   & $\frac{1}{6}\sum_{ijk=1}^3\epsilon_{ijk}[\gamma_i,\gamma_j,\gamma_k]$ & $3/2$ & $0$ \cr
 III  & $\frac{1}{3}\sum_{k=1}^3[\gamma_5,\gamma_k,\gamma_k]$      & $3/2$ & $0$ \cr 
 IV   & $\frac{1}{3}\sum_{k=1}^3[\gamma_k,\gamma_5,\gamma_k]$      & $3/2$ & $0$ \cr 
 V    & $\frac{1}{3}\sum_{k=1}^3[\gamma_k,\gamma_k,\gamma_5]$      & $3/2$ & $0$ \cr 
 VI   & $\frac{1}{6}\sum_{ijk=1}^3\epsilon_{ijk}[\gamma_i,\gamma_j,\gamma_k]$ & $0$ & $3/2$ \cr
 VII  & $\frac{1}{3}\sum_{k=1}^3[\gamma_5,\gamma_k,\gamma_k]$      & $1/2$ & $1$ \cr 
 VIII & $\frac{1}{3}\sum_{k=1}^3[\gamma_k,\gamma_5,\gamma_k]$      & $1/2$ & $1$ \cr 
 IX   & $\frac{1}{3}\sum_{k=1}^3[\gamma_k,\gamma_k,\gamma_5]$      & $1/2$ & $1$ \cr 

 \hline\\[-1.8ex]
  \\
 \end{tabular}
\caption{}
 \label{tab:bk}
\vspace{-8mm}
\end{table}

\section{One--loop perturbative expansion}

In order to extract the one--loop contribution to the renormalization
constant, all the correlation functions must be expanded in powers of
the bare coupling,
\be
{\cal X} = {\cal X}^{(0)} + g_0^2\biggl[{\cal X}^{(1)} +
  m_c^{(1)}\frac{\partial {\cal X}^{(0)}}{\partial
    m_0}\biggr] + O( g_0^4),
\ee
with ${\cal X}$ being $h^\pm$, $f_1$, $k_1$, as well as the renormalization constant itself,
\be
Z^{\pm}_{\rm VA+AV} = 1 + g_0^2 Z^{\pm(1)}_{\rm VA+AV} + O( g_0^4).
\ee
Terms proportional to the one--loop coefficient of the critical mass
$m_c^{(1)}$ are needed to set the correlation functions to zero
renormalized quark mass. The values of $f_1^{(0,1)}$ and
$k_1^{(0,1)}$ are known from the literature \cite{Sint:1997jx}, while
$h^{\pm(0,1)}$ have been computed through numerical integration of the related Feynman
diagrams for both Wilson and Clover actions. 
\vskip 0.2cm
The analytic structure of $Z_{\rm VA+AV}^{\pm(1)}$ is provided by the 
``log''--divergent relation
\be
Z_{\rm VA+AV}^{\pm(1)} = B^{\pm}_{\rm SF} +
\gamma^{\pm(0)}\ln\left(\frac{a}{L}\right) + O\left(\frac{a}{L}\right),
\ee
where $\gamma^{\pm(0)}=\pm 6(1 \mp 1/N)/(4\pi)^2$ is the LO anomalous dimension of
${\cal O}_{\rm VA+AV}^\pm$
and $B^\pm_{\rm SF}$ is a finite term, which depends upon the renormalization
scheme and can be computed by inserting the perturbative expansion
of the correlation functions and the renormalization constant in eq.~(7).

\section{Matching to DRED}

Given the NLO (scheme dependent) anomalous dimension in a
reference scheme, it can be computed in any other scheme, provided the 
one--loop order renormalization constant is known in the latter. In
our case we choose DRED as the reference scheme, and the matching formula
is given by
\be
\gamma_{\rm SF}^{\pm(1)} = \gamma_{\rm DRED}^{\pm(1)} +
2b_0\biggl[B^\pm_{\rm SF} -
      {z^\pm}\biggr] +
\gamma^{\pm(0)}{c_g},
\ee
where the second term on the RHS measures the shift of the finite term
of the one--loop renormalization constant and the third one
measures the shift of the coupling between the schemes. The
coefficients $\gamma_{\rm DRED}^{\pm(1)}$, $z^\pm$ and $c_g$ can be
retrieved from the literature
\cite{Altarelli:1980fi,Martinelli:1983ac,Frezzotti:1991pe,Sint:1995ch} and
$b_0$ is the LO contribution to the $\beta$--function. In Table~2 we
report the (preliminary) values of
$\gamma^{\pm(1)}$ in the schemes defined in Table~1. The
dependence of the NLO--AD upon $\rm N_f$ is implicit in the
coefficients $\gamma^{\pm(1)}_{\rm DRED}$, $b_0$ and $c_1$.
\begin{table}[t]
{
 \centering
 \begin{tabular}{cc}
 \hline\\[-1.8ex]
 scheme & $\gamma^{+(1)}\times 10^{3}$ \\
 \hline\\[-1.8ex]
 I    & $\ \ 0.517(3) + 0.2032(2) \ {\rm N_f}$ \cr
 II   & $-6.072(5) + 0.6026(3) \ {\rm N_f}$ \cr
 III  & $\ \ 1.690(3) + 0.1321(2) \ {\rm N_f}$ \cr
 IV  & $-7.966(5) + 0.7174(3) \ {\rm N_f}$ \cr
 V    & $-7.290(5) + 0.6763(3) \ {\rm N_f}$ \cr
 VI   & $-7.830(8) + 0.7090(5) \ {\rm N_f}$ \cr
 VII  & $\ \ 0.519(5) + 0.2031(3) \ {\rm N_f}$ \cr
 VIII  & $-9.139(8) + 0.7885(5) \ {\rm N_f}$ \cr
 IX   & $-8.463(8) + 0.7475(5) \ {\rm N_f}$ \cr
 \hline\\[-1.8ex]
 scheme & $\gamma^{-(1)}\times 10^2$ \\
 \hline\\[-1.8ex]
 I    & $2.3660(3) - 0.19711(2) \ {\rm N_f}$  \cr
 II   & $0.9320(5) - 0.11020(3) \ {\rm N_f}$  \cr
 III   & $2.4833(3) - 0.20422(2) \ {\rm N_f}$  \cr
 IV   & $0.5530(4) - 0.08723(3) \ {\rm N_f}$  \cr
 V    & $0.8057(4) - 0.10255(3) \ {\rm N_f}$  \cr
 VI   & $0.7559(5) - 0.09953(3) \ {\rm N_f}$  \cr
 VII & $2.3660(3) - 0.19711(2) \ {\rm N_f}$  \cr
 VIII  & $0.4355(4) - 0.08011(3) \ {\rm N_f}$  \cr
 IX   & $0.6882(4) - 0.09543(3) \ {\rm N_f}$  \cr

 \hline\\[-1.8ex]
  \\
 \end{tabular}
\caption{}
 \label{tab:bk2}}
\vspace{-8mm}
\end{table}

\section{NP vs. NLO running of ${\cal O}_{\rm VA+AV}^\pm$}

The knowledge of the NLO anomalous dimension allows, in particular, to compare the
nonperturbative and NLO (scheme dependent) running. To do this, we consider
the step scaling function (ssf)
\be
\sigma^\pm_{\rm VA+AV}(u) = \lim_{a\to
  0}\frac{Z^\pm_{\rm VA+AV}(g_0,a/2L)}{Z^\pm_{\rm VA+AV}(g_0,a/L)}\biggl|_{u=\bar g^2},
\ee
which has been computed nonperturbatively in \cite{Guagnelli:2002rw} and expand it to
NLO:
\[
\sigma^\pm_{\rm VA+AV}(u) = 1 + \sigma_1^\pm u + \sigma_2^\pm u +
O(u^3)
\]\vskip -0.4cm
\begin{eqnarray}
\sigma^\pm_1 & = & \gamma^{\pm(0)}\ln2,\cr
\sigma^\pm_2 & = & \gamma^{\pm(1)}\ln2\ + \cr
&  & \bigl[\frac{1}{2}(\gamma^{\pm(0)})^2 +b_0\gamma^{\pm(0)}\bigr](\ln
2)^2.
\end{eqnarray}
As an example, in Figure~1 we report plots of the ssf $\sigma^+_{\rm
VA+AV}$ in the schemes I and II. In both plots, the solid line
represents the LO running, the dashed line represents the NLO 
running and points are the nonperturbative
data. The agreement between perturbation theory
and nonperturbative data is scheme dependent; in some schemes 
the agreement improves when switching to the NLO, while in some others
it worsens. 

\begin{figure}[h]
\vskip -0.5cm
\includegraphics[width=3.7cm]{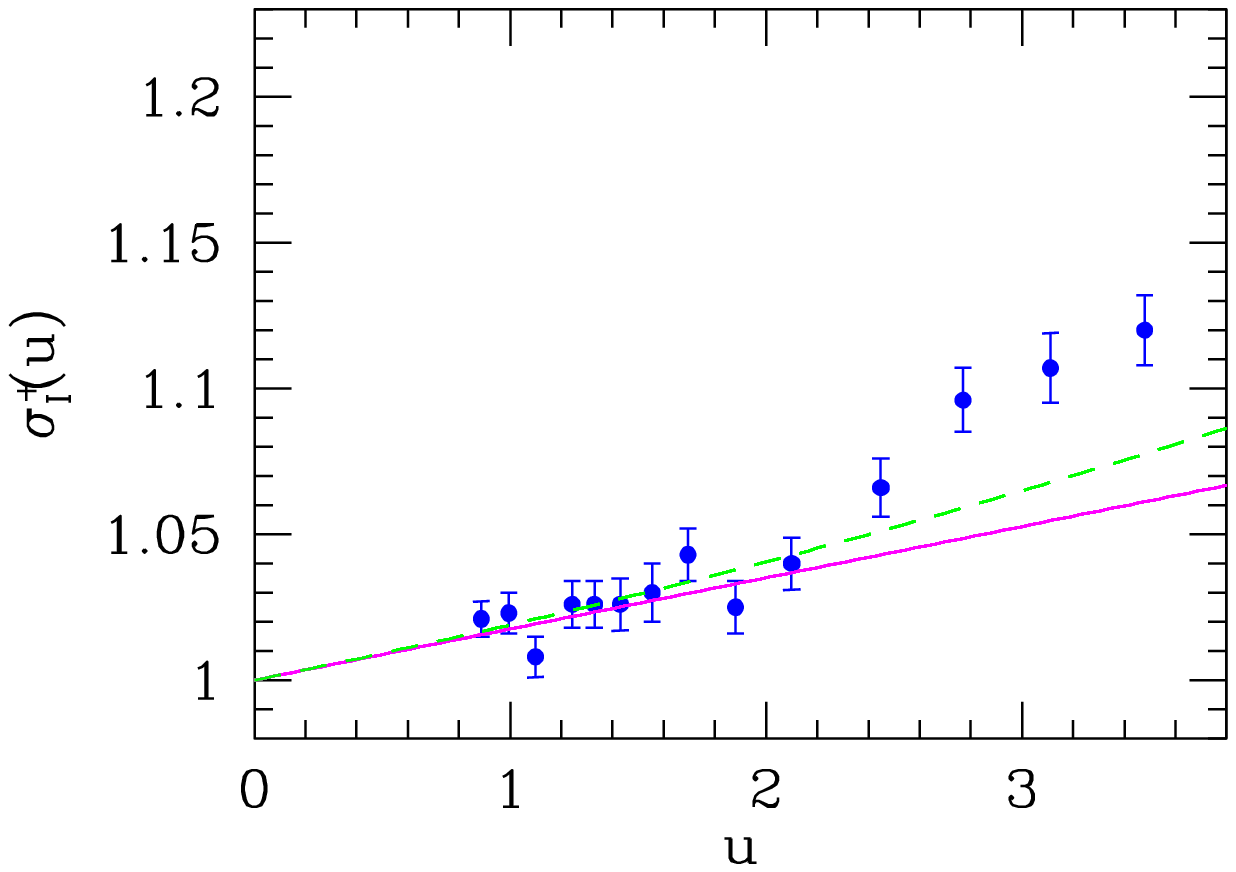}
\vskip -2.6cm
\hskip 3.9cm\includegraphics[width=3.7cm]{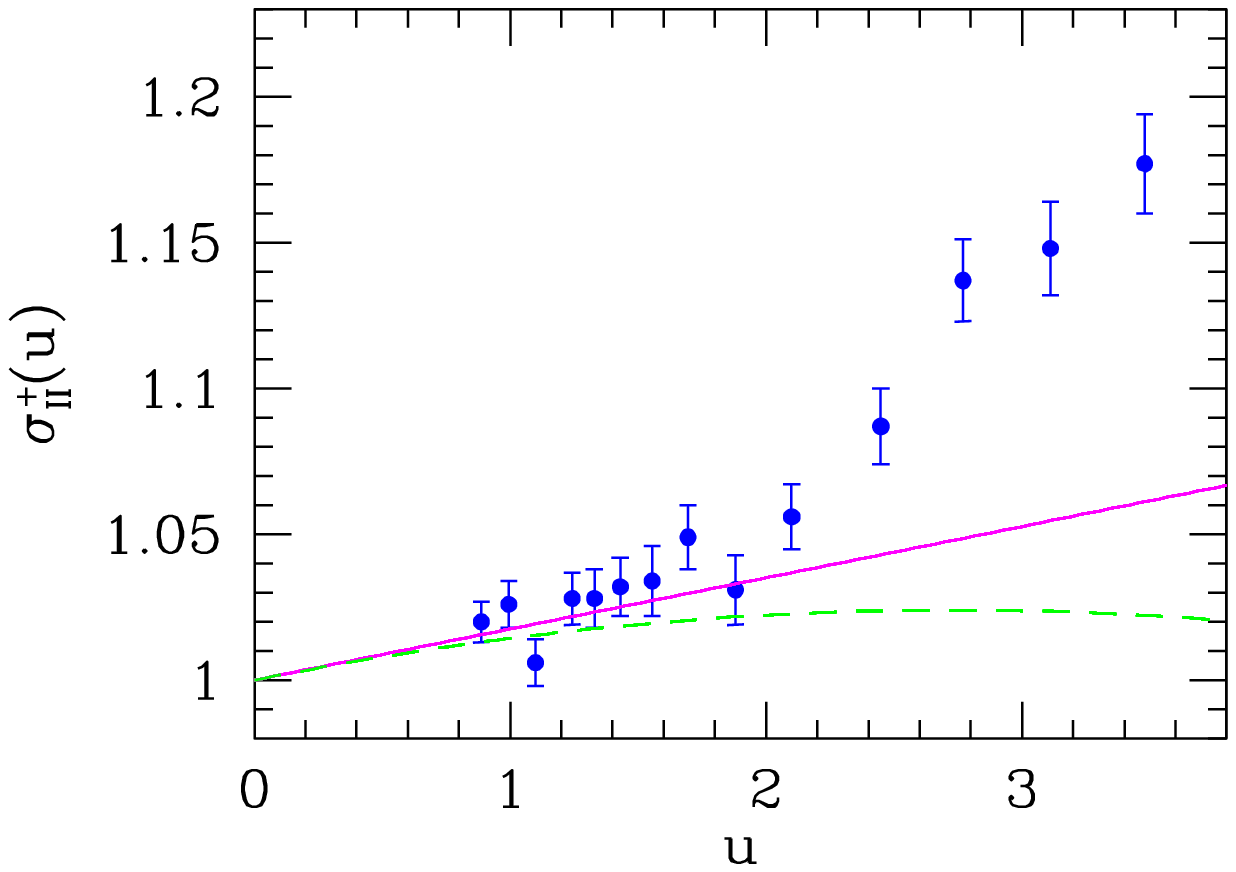}
\vskip -0.8cm
\caption{\small $\sigma^+_{\rm VA+AV}(u)$ (schemes I and II): 
LO running (continuous line), NLO running (dashed line) and NP data
(points) from \cite{Guagnelli:2002rw}.}
\label{fig:6737}
\vskip -0.9cm
\end{figure}
We conclude that the SF scheme has to be chosen with care in order to
achieve a good control of the matching to the continuum schemes.



\end{document}